\documentclass[11pt]{article}
\usepackage{graphicx}

\newcommand{\BABARPubYear}    {02}

\newcommand{\BABARConfNumber} {08}
\newcommand{\SLACPubNumber} {9231}
\newcommand{\LANLNumber} {0000}

\input pubboard/babarsym

\def\bdspi    {\ensuremath {\Bz \to D_s^{+} \pi^-}}
\def\bdsspi    {\ensuremath {\Bz \to D_s^{*+} \pi^-}}

\def\bdsordsspi   {\ensuremath {\Bz \to D_s^{(*)+}\pi^-}}

\def\sin2bg       {\ensuremath {\sin (2\beta+\gamma)}}
\def\btou       {\ensuremath {b\rightarrow u}}
     
\def\brdsShort {\ensuremath { 3.1 \pm 1.0\, ({\rm stat. }) }}
\def\brdssShort {\ensuremath { 2.1^{+0.8}_{-1.0}\, ({\rm stat. }) }}

\def\brds {\BR(\bdspi)~=~(\brdsShort $\pm 1.0\, ({ \rm syst. })$) \ensuremath{\times 10^{-5}}}
\def\brdss {\BR(\bdsspi)~=~(\brdssShort$\pm 0.6\, ({ \rm syst. })$) \ensuremath{\times 10^{-5}}}
\def\brdsNoSys {\BR(\bdspi)~=~(\brdsShort) \ensuremath{\times 10^{-5}}}
\def\brdssNoSys {\BR(\bdsspi)~=~(\brdssShort) \ensuremath{\times 10^{-5}}}
\def\brdsslim {\BR(\bdsspi)\ensuremath{~<~4.3\times 10^{-5}}~at~90\%~C.L.}

\def\lumi    {\ensuremath {61.6\times 10^6}}

\def\cth     {\ensuremath {|\cos{\theta_{T}}|}}
\def\cthel     {\ensuremath {|\cos{\theta_{H}}|}}

\def\ebeam     {\ensuremath {E^{*}_{beam}}}
\def\esqbeam     {\ensuremath {E^{*2}_{beam}}}

\def\de        {\ensuremath {\Delta E^{*}}}

\def\mpl #1 #2 #3 {Mod.~Phys.~Lett.~{\bf#1},\ #2 (#3)}
\def\npb  #1 #2 #3 {Nucl.~Phys.~B~{\bf#1},\ #2 (#3)}
\def\plb  #1 #2 #3 {Phys.~Lett.~B~{\bf#1},\ #2 (#3)}
\def\pr   #1 #2 #3 {Phys.~Rep.~{\bf#1},\ #2 (#3)}
\def\prd  #1 #2 #3 {Phys.~Rev.~D~{\bf#1},\ #2 (#3)}
\def\prl  #1 #2 #3 {Phys.~Rev.~Lett.~{\bf#1},\ #2 (#3)}
\def\RMP  #1 #2 #3 {Rev.~Mod.~Phys.~{\bf#1},\ #2 (#3)}
\def\zpc  #1 #2 #3 {Z.~Phys.~C~{\bf#1},\ #2 (#3)}
\def\nim  #1 #2 #3 {Nucl.~Instrum.~Methods~{\bf#1},\ #2 (#3)}
\def\nima  #1 #2 #3 {Nucl.~Instrum.~Methods~A.{\bf#1},\ #2 (#3)}
\def\epjc #1 #2 #3 {Euro.~Phys.~Jour~C~{\bf#1},\ #2 (#3)}
\def\rmp #1 #2 #3 {Rev.~Mod.~Phys~{\bf#1},\ #2 (#3)}
\def\npbps #1 #2 #3 {Nucl.~Phys.~B.~proc.~suppl~{\bf#1},\ #2 (#3)}
\def\progtp #1 #2 #3 {Prog.~Theo.~Phys~{\bf#1},\ #2 (#3)}
\def\etal{{\it et al.}}

\long\def\inst#1{\par\nobreak\kern 4pt\nobreak
    {\it #1}\par\vskip 10pt plus 3pt minus 3pt}

\begin{document}
{\pagestyle{empty}

\begin{flushright}
\babar-CONF-\BABARPubYear/\BABARConfNumber \\
SLAC-PUB-\SLACPubNumber \\
hep-ex/\LANLNumber \\
 
\end{flushright}

\par\vskip 5cm

\begin{center}
\Large \bf Evidence for the \btou\ transition   \bdspi\ and a search for 
\bdsspi
\end{center}
\bigskip

\begin{center}
\large The \babar\ Collaboration\\
\mbox{ }\\
\today
\end{center}
\bigskip \bigskip

\begin{center}
\large \bf Abstract
\end{center}
We report  evidence for  the \btou\ transition \bdspi\ 
 and the results of the search for \bdsspi\ 
from a sample  of \lumi\ \FourS\ decays into $B$ meson pairs collected with the \babar\ detector
at the PEP II asymmetric $e^{+}e^{-}$ collider. 
 The observed \bdspi\ yield has a probability of $4.4 \times 10^{-4}$ to be a fluctuation of the background (3.5 $\sigma$)
and we measure the branching fraction \brds. We also set a limit \brdsslim

\vfill
\begin{center}
Presented at the Flavor Physics and CP Violation Conference,\\ 
5/16---5/18/2002, Philadelphia, PA
\end{center}

\vspace{1.0cm}
\begin{center}
{\em Stanford Linear Accelerator Center, Stanford University, 
Stanford, CA 94309} \\ \vspace{0.1cm}\hrule\vspace{0.1cm}
Work supported in part by Department of Energy contract DE-AC03-76SF00515.
\end{center}

\newpage
} 

\begin{center}
\small

The \babar\ Collaboration,
\bigskip

B.~Aubert,
D.~Boutigny,
J.-M.~Gaillard,
A.~Hicheur,
Y.~Karyotakis,
J.~P.~Lees,
P.~Robbe,
V.~Tisserand,
A.~Zghiche
\inst{Laboratoire de Physique des Particules, F-74941 Annecy-le-Vieux, France }
A.~Palano,
A.~Pompili
\inst{Universit\`a di Bari, Dipartimento di Fisica and INFN, I-70126 Bari, Italy }
% ...

\end{center}\newpage



The measurement of the \CP-violating phase of the Cabibbo-Kobayashi-Maskawa  (CKM) matrix \cite{CKM}
is an important part of the present scientific program in particle physics. 
\CP\ violation manifests itself as a non-zero area of the unitarity triangle \cite{Jarlskog}:
while it is sufficient to measure one of the angles to demonstrate the existence of \CP\ violation,
more than one angle must be measured to demonstrate that the CKM mechanism is the correct 
explanation of this phenomenon. There are already several theoretically clean measurements
of the angle $\beta$~\cite{sin2b} but there is no such measurement of the other two angles ($\alpha$ and 
$\gamma $). A theoretically clean measurement of $\rm sin(2\beta+\gamma)$ can be obtained from the study
of the time evolution of the $\Bz\to D^{(*)+}\pi^-$ and  $\Bz\to D^{(*)-} \pi^+$\cite{sin2bg} decays.
 This measurement 
requires a knowledge of the ratio 
between the decay amplitudes of these two processes. 
Unfortunately $\Bz\to D^{(*)+}\pi^-$  decays, which are Cabibbo suppressed,
 cannot be efficiently isolated from 
 $\Bzb\to D^{(*)+}\pi^-$ decays.
Since the amplitude of the decay  
$\Bz\to D^{(*)+} \pi^-$ can be related to the decay \bdsordsspi  \cite{sin2bg}, 
the measurement of  $\rm sin (2\beta+\gamma)$ will require the knowledge
of \BR($\bdsordsspi$).
Decays of this type have also been proposed to be used for the measurement of   $ |V_{ub}/V_{cb}|$ \cite{roy}.
In this paper we present the results of the measurement of the
branching fractions of the decays  \bdsordsspi.  


The data were collected  in the years 2000-2001 with the \babar\ detector
at the \pep2\ asymmetric $\ep (3.1\gev ) $ -- $\en (9\gev ) $  storage ring~\cite{pep}.
Since the \babar\ detector is described in detail
elsewhere~\cite{detector},  
only a brief description is given here. 
Surrounding the beam-pipe is a Silicon Vertex Tracker (SVT), which
provides precise measurements of the trajectories of charged particles
as they leave the \epem interaction point.  
Outside of the SVT, a 40-layer drift chamber (DCH) 
allows measurements of track momenta in a 1.5\,T magnetic field as well as
energy-loss measurements, which contribute to charged
particle identification. Surrounding the DCH is a Detector of Internally
Reflected Cherenkov radiation (DIRC), which provides charged hadron
identification. Outside of the DIRC is a CsI(Tl) electromagnetic
calorimeter (EMC) that is used to detect photons, provide electron
identification and reconstruct neutral hadrons. The EMC is surrounded by a
superconducting coil, which creates the magnetic field for momentum
measurements.  Outside of the coil, the 
flux return is instrumented with resistive plate chambers interspersed with 
iron (IFR) for the identification of muons and long-lived neutral hadrons.
We use the GEANT~\cite{geant} software to simulate interactions of particles
traversing the \babar\ detector, taking into account the varying detector conditions and beam backgrounds.

We select hadronic events with a minimum of three reconstructed charged tracks having  an impact parameter in
the plane transverse to the beam less than 0.5\cm\ from the beam-line.
The event must have a total measured energy in the laboratory greater than 4.5\gev 
within the fiducial regions for charged tracks and neutral clusters with energy above 30\mev. 
To help reject continuum background, the ratio of the second-to-zeroth Fox-Wolfram moment~\cite{fox} 
must be less than 0.5.

Only upper limits on these modes have been reported~\cite{prior} and therefore 
the selection criteria are optimized, prior to looking 
at the signal yield in data, to maximize $\epsilon_S/\sqrt{n_B}$. The signal efficiency
$\epsilon_S$ is evaluated from Monte Carlo,
while the background $n_B$ is evaluated with data sidebands and confirmed by Monte Carlo simulation. 
The criteria are then made uniform among decay modes when appropriate.

The decays \bdsordsspi\ 
are reconstructed in the modes
 $\Dss\to\Ds\gamma$, $\Ds \to \phi \pi^+$, $\KS K^+$ and $\Kstarzb\Kp$, 
with $\phi\to K^+K^-$, $\KS \to \pip \pim$ and $\Kstarzb\to K^-\pi^+$. 
The  $\Kp$ and $\pim$ track candidates  are 
required 
 to originate from a vertex consistent with the
$\epem$ interaction point. The $\KS$ candidates are reconstructed from two
oppositely-charged tracks with an invariant mass 
$493 < M_{\pip \pim} < 501\mevcc $.
The $\phi$ candidates are reconstructed from two oppositely-charged 
kaons with an invariant mass $1009 < M_{\Kp \Km} < 1029\mevcc $. 
A track is identified as a kaon using the information
from the energy-loss measurement in the SVT
(only for tracks for momentum $p<0.5\gevc$), the DCH (if $p\le 0.6\gevc$) and the
measured 
Cherenkov angle in the DIRC
(if $p>0.6\gevc$). Two likelihood selections based on these quantities are used in this analysis: either very
 loose criteria with 95\% efficiency and 20\% 
pion contamination, or tight criteria with 85\% efficiency and 5\% pion contamination. If not otherwise specified the
former is adopted.
 To reconstruct $\Kstarz$ candidates, a \Kp\ and a \pim\ are required to have an invariant mass 
$856 < M_{\Kp \pim} < 936\mevcc $.
The polarization of the 
 \Kstarz\ and $\phi$ mesons in the   \Ds\ decays is also utilized to reject backgrounds through the use of the helicity angle
$\theta_H$, defined as the angle between one of the decay products 
and the direction of flight of the meson, in the meson rest frame. Background events are distributed uniformly in $\cthel$, while
signal events are distributed as $\cthel^2$.
\Kstarz\ candidates are therefore required to have $\cthel>0.4$, while
for the $\phi$ candidates we require $\cthel>0.5$. 
In order to reject background from \Dp\to\KS\pip or \Kstarz\pip,
the additional kaon in the reconstruction of \Ds\to\KS\Kp or \Kstarzb\Kp is required to pass the tight criteria.
Finally, the \Ds\ candidates are required to have an invariant mass within 10 \mevcc\ 
of the nominal mass~\cite{PDG2000}.

 \Dss\ candidates are reconstructed by combining \Ds\ and photon candidates. 
The \Ds\ selection is re-optimized for the \bdsspi\ case, but the resulting
selection criteria are very similar to that of the \bdspi\ case.
Photons that form a $\pi^0$ candidate in combination  with
any other photon  with energy greater than 70 \mev are not considered.
 The mass difference between the \Dss\ and the the \Ds\ candidate is required to be within 14 \mevcc\ of the nominal value~\cite{PDG2000}. 


\begin{figure}[!htb]
\begin{center}
\includegraphics[width=0.5\linewidth]{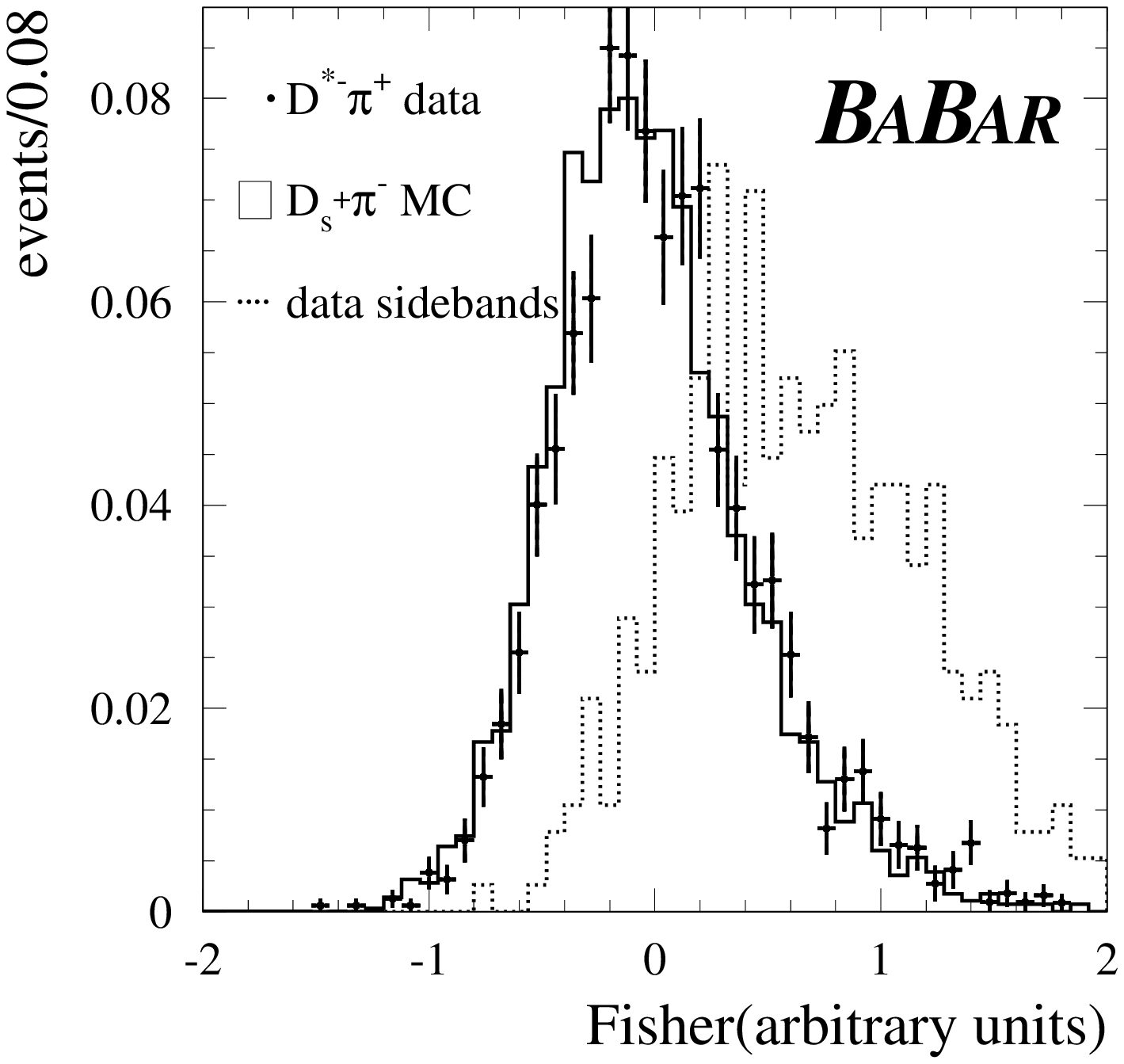}
\caption{Distribution of the Fisher discriminant  $\cal{F}$ in \bdspi\ Monte Carlo, $\Bz\to\Dstarm\pip$ data, which  
has the same form as the signal, and 
\de\ sideband data, which  
has the same form as the background. }
\label{fig:costhr}
\end{center}
\end{figure}

$B$ meson candidates are reconstructed from the \Ds\ or \Dss\ candidates and a charged track,
which is required to fail the 
tight kaon criteria. 
Since the $B$ mesons are produced via $\epem \to \Y4S \to \BB$, the energy of
the $B$ meson in the center-of-mass frame is the beam energy, {$\ebeam$}.
The distribution of $\de=E^{*}_{B} - \ebeam$ peaks
 at zero for the signal. The 
  \de\ signal band is defined by 
$|\de|<36 \mev$ 
 while the \de\  sidebands
are defined as  the remaining events with $|\de|<120 \mev$.

Backgrounds coming from other $B$ decays, such as $\Bz\to\Dm\pip,\rho^+$ with $\Dm\to\KS\pim$ or $\Kstarz\pim$ 
deserve particular attention because their reconstructed $B$ mass has a shape very similar to that of the signal
(peaking background). In these events one of the pions in the \Dm\ decay is misidentified as a kaon and the distribution of the reconstructed 
invariant mass of the \Dm\ decay products overlaps with the signal \Ds\ mass distribution.
Nonetheless, since their \de\ distribution does not peak at  $\de=0$, their contribution in the signal region is small.
Modes with the same final state as the signal but with no intermediate \Ds\ can also constitute a peaking background. They have the same 
\de\ distribution as the signal, but not the same distribution of the invariant mass for the \Ds\ candidate.

In order to reject events where the \Ds
comes from a $B$ candidate and the pion from the other $B$, we require the two candidates to
have a probability greater than 0.25\% of originating from the same vertex, which is 98\% efficient on the signal.
The remaining background is predominantly from continuum \qqbar\ production with a \Ds, a $\phi$ or a
\Kstarz\ meson produced in the hadronization of one of the two quarks.
We use  event topology differences between signal
and background to reduce the continuum contribution. We compute one thrust axis  
using only the  $B$ meson decay product candidates and one with all the other tracks.
The angle between the two thrust axes ($\theta_T$) is used to discriminate the background.  
In the center-of-mass frame, \BB\ pairs are
produced approximately at rest and  produce a uniform $\cth$ distribution.
In contrast,
\qqbar\ pairs are produced back-to-back in the center-of-mass frame, which results in a $\cth$ distribution peaking at
1. 
Depending on the background level of each mode we require either $\cth < 0.8$ or $<0.7$.
 We further suppress backgrounds using a Fisher discriminant $\cal{F}$ constructed from
the scalar sum of the center-of-mass momenta of all tracks and photons (excluding the $B$ candidate
decay products)  flowing into 9 concentric cones centered on the thrust axis of the $B$ candidate
\cite{twobody}. The more spherical the
event, the lower the value of ${\cal{F}}$. Figure~\ref{fig:costhr} shows the distribution of this variable 
in data sidebands, which have the characteristics of  the background, 
in simulated signal events and in a control sample of approximately 1500 $\Bz\to\Dstarm\pip$ fully reconstructed events
with $\Dstarm\to\Dzb\pim$ and $\Dzb\to\Kp\pim$. It is a copius decay channel,
 with low background, similar final state, therefore well suited to
 investigate signal properties in data.
The signal distribution is well reproduced by the simulation.
 We require ${\cal{F}}<0.05$ for the $\Ds\to\phi\pip$ and \KS\Kp
modes and ${\cal{F}}<0.2$ for  \Ds\to\Kstarzb\Kp . 
The selection criteria for \bdsspi\ are optimized separately, but are  close to the ones for \bdspi.
 

As a measure of the $B$ meson mass, the beam-energy substituted mass is  defined as 
$\mes= \sqrt{ E^{*2}_{beam}-\mbox{\boldmath $\mathrm p$}_{B}^{*2}}$, where \mbox{\boldmath $\mathrm p$}$^{*}_{B}$ 
is the momentum vector of the $B$ meson candidate in the center-of-mass frame, calculated from the measured 
momenta of the decay products. The $\mes$ distribution for the signal is well
described by a Gaussian distribution  dominated by the resolution of
the beam energy measurement, and therefore independent of the decay mode.
The combinatorial background is empirically 
described by a threshold function (the so called ``ARGUS'' shape
$
dN/d\mes =A_B \mes \sqrt{1-\mes^{2}/\esqbeam}
             \exp\left[-\zeta\left(1-\mes^{2}/\esqbeam\right)\right],
$
a function introduced by the  ARGUS Collaboration~\cite{argusf} 
) for each mode.

\begin{figure}[!htb]
\begin{center}
\includegraphics[width=0.45\linewidth]{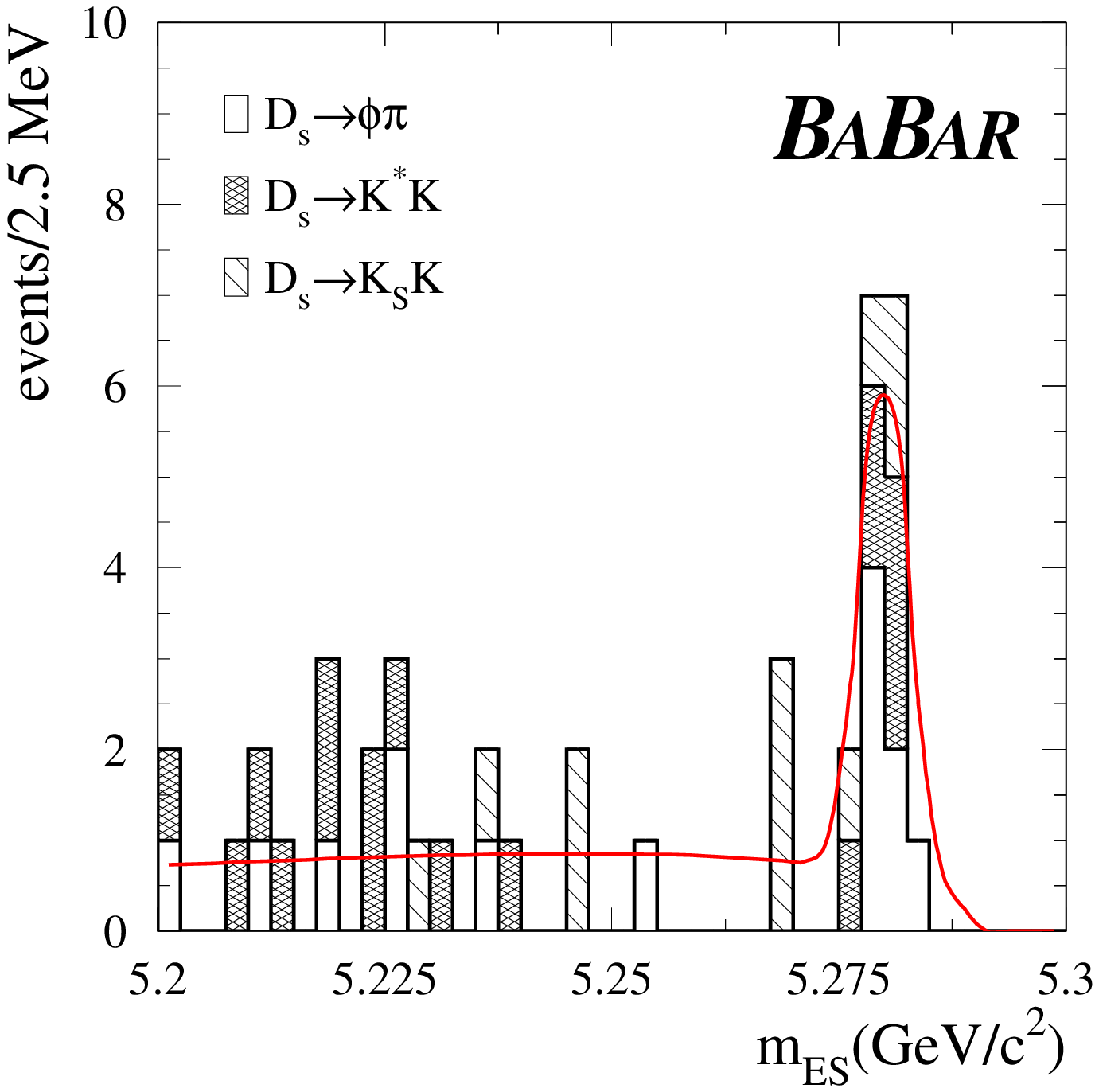}
\includegraphics[width=0.45\linewidth]{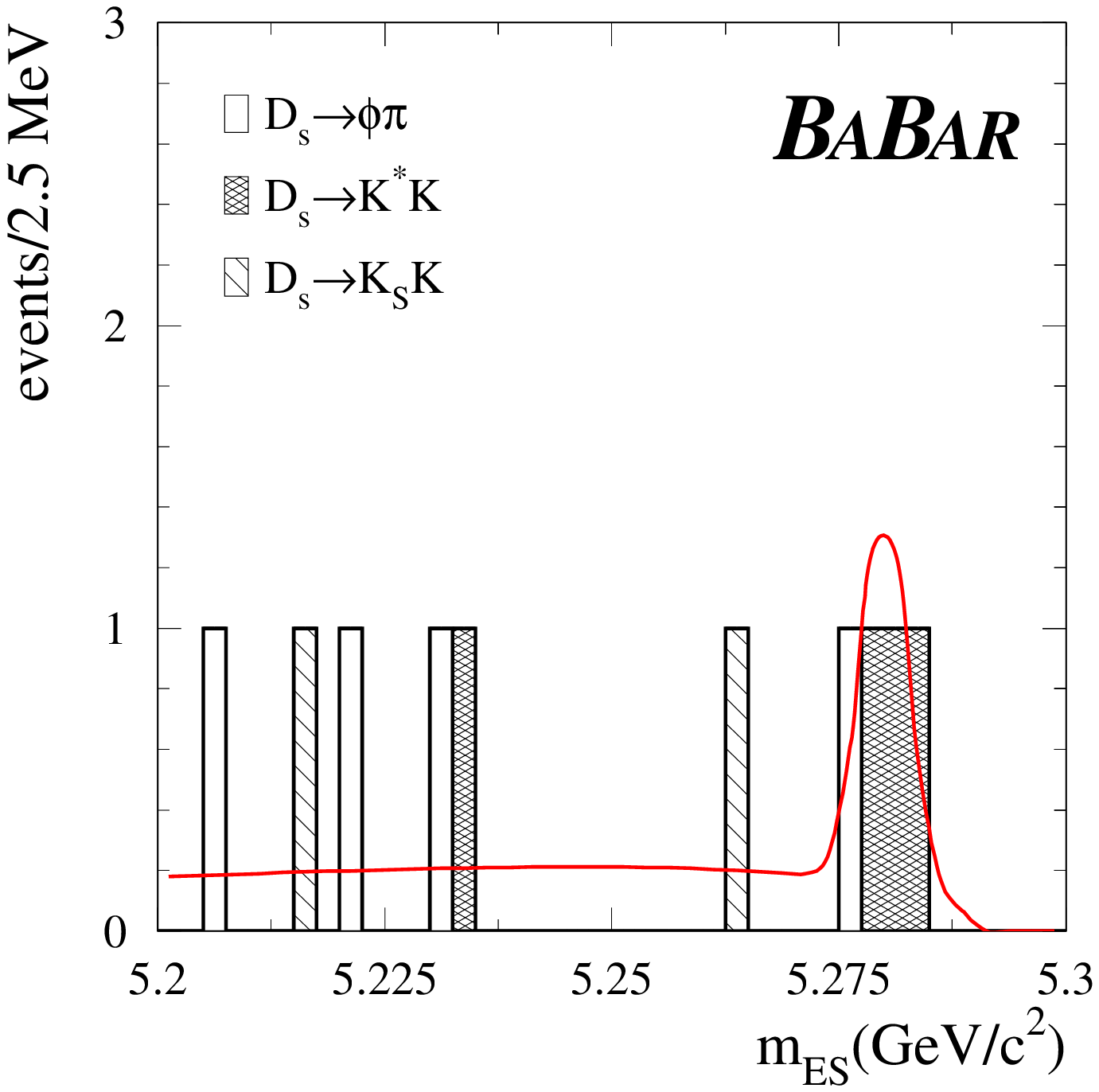}
\caption{Distribution of \mes\ for the  \bdspi\ (left) and \bdsspi\ (right)
candidates after all selection criteria, including the \de\ signal window. 
The fits used to obtain the signal yield are described in the text. The contributions from the
individual modes are also shown. 
}
\label{fig:datamb}
\end{center}
\end{figure}

Figure~\ref{fig:datamb} shows the \mes\ distribution for each of the modes. 
An unbinned maximum-likelihood fit 
is used to fit the \mes\ distributions
for signal and ARGUS shaped background contributions. The different \Ds\ decay modes are
combined. The mean and width of the signal distribution are fixed to the 
values obtained in a high statistics $\Bz\to D^{(*)-} \pip$ sample.
The ARGUS shape parameter~\cite{argusf} is fixed to the value  fitted on the data 
after having released, in order to increase the statistics, the requirements on the 
\Ds\ mass (accepted within 40\mevcc of the nominal mass) and the \de (to 50\mev).
The signal yields and combinatorial background, as returned from the fit,
 are given in Table~\ref{tab:fit}.   


\begin{table}[!htb]
\caption{The number of events in the signal box ($N_{sigbox}$ ), 
the  signal yield ($N_{sig}$) and  the combinatorial background ($N_{comb}$) as extracted from the likelihood fit, 
the efficiency ($\varepsilon$), the peaking background ($N_{peak}$), and
the measured branching fraction (\BR ). $N_{sig}$ and $N_{comb}$ and \BR\  are not available for modes with too few events.
$N_{peak}$ is not reported if no event is found in the \Ds\ mass sideband. Only statistical errors are quoted. }
\begin{center}
\mbox{ \scriptsize
\begin{tabular}{ccccccr} \hline \hline
 \ \ \ \Ds\ Mode \ \ \ \ \ \ \     & \ \ \  \  $\varepsilon$  \ \ \ \ \ \    & \ \ \  \  $N_{sigbox}$  \ \ \ \ \ \    &\ \  \ \ \  $N_{sig}$ \ \ \ \ \ \ \  & \ \  \ \ \  $N_{comb}$ \ \ \ \ \ \ \     &   \ \ \  $N_{peak}$    \ \ \ \ \ \ \        &  \ \ \ \ \ \ $\BR $   \ \ \ \  \\
            & \% &  events  & events  &  events    &  events            & $10^{-5}$\\ \hline
\multicolumn{5}{c}{\bdspi}\\ \hline 
$\phi\pip$     &  16.6    & 7 & 6.4 $\pm$ 2.7& 1.2$\pm$0.5&  -    & $ 3.5 \pm 1.5 $   \\ 
$ \Kstarzb\Kp$       & 9.7 & 6    & 5.1$\pm$2.4 & 1.9$\pm$0.6&$2.3 \pm 1.8$  & $2.2 \pm 1.8 $ \\
$\KS\Kp    $   &   12.2  & 4 & 3.4$\pm$2.0&1.2$\pm$0.5 & -  & $3.7 \pm 2.2  $ \\
total          &    & 17   &14.9$\pm$4.1 &  4.4$\pm0.9$& $2.3 \pm 1.8$  & 3.1 $ \pm  1.0$ \\\hline
\multicolumn{5}{c}{\bdsspi}\\ \hline 
$\phi\pip$     &    7.8  & 1  &   &   & - &   \\ 
$\Kstarzb\Kp$       &   $3.3$   & 3 & $2.9\pm1.8$& 0.17$\pm$0.17 & 0.20 $\pm$ 0.14  & $6.5^{+3.6}_{-2.6}$ \\ 
$\KS\Kp$       & 5.1    & 0 &  &   &  - &\\ \hline
total          &   & 4   & $3.5\pm 2.0$&  0.94$\pm$0.38 & 0.20 $\pm$ 0.14 &  $2.1^{+0.8}_{-1.0}$  \\\hline
\end{tabular}}
\end{center}
\label{tab:fit}
\end{table}


The efficiency for the selection requirements is given in Table~\ref{tab:fit} as obtained from  simulation.
The measurement of the branching fractions for the individual modes are shown as a cross check
 but they are not used to obtain the result. 
 The branching fraction is determined from the signal yield  ($N_{sig}$), 
 the peaking background ($N_{peak}$), the efficiency ($\varepsilon$)
and the total number of \BB\ events in the sample ($N_{BB}=61.6\pm 0.6 \times  10^6$). 
The fit separates the combinatoric background ($N_{comb}$) from
$N_{sig}$ which is the sum of the signal and the peaking
background.
We estimate the total peaking background by fitting the \mes\ distribution in the \Ds\ mass sidebands.
The observed yield is rescaled to the signal region based on the \Ds\ mass distribution in background events. 
The only peaking background that would not be included in this estimate comes from $B$ decays with a \Ds\ in the final state. 
Simulation of a large number of events in these decay modes shows their contribution to be negligible.

Within a 3$\sigma$  \mes\ window we find 17 \bdspi\ and 4 \bdsspi\ events, and a gaussian component 
(signal and peaking background) of 14.9$\pm$4.1 and 
3.5$\pm$2.0 events, respectively. The \bdspi\ yield has a probability of  $4.4\times 10^{-4}$ to be a fluctuation of the
 background 
($3.5\sigma$) and we measure a branching fraction \brdsNoSys. 
 The \bdsspi\ yield has a probability of $2.2\sigma$ to be a background fluctuation
 and the measured branching fraction is \brdssNoSys. 
Probabilities are computed including all the uncertainties on the backgrounds.

\begin{figure}[!htb]
\begin{center}
\includegraphics[width=0.45\linewidth]{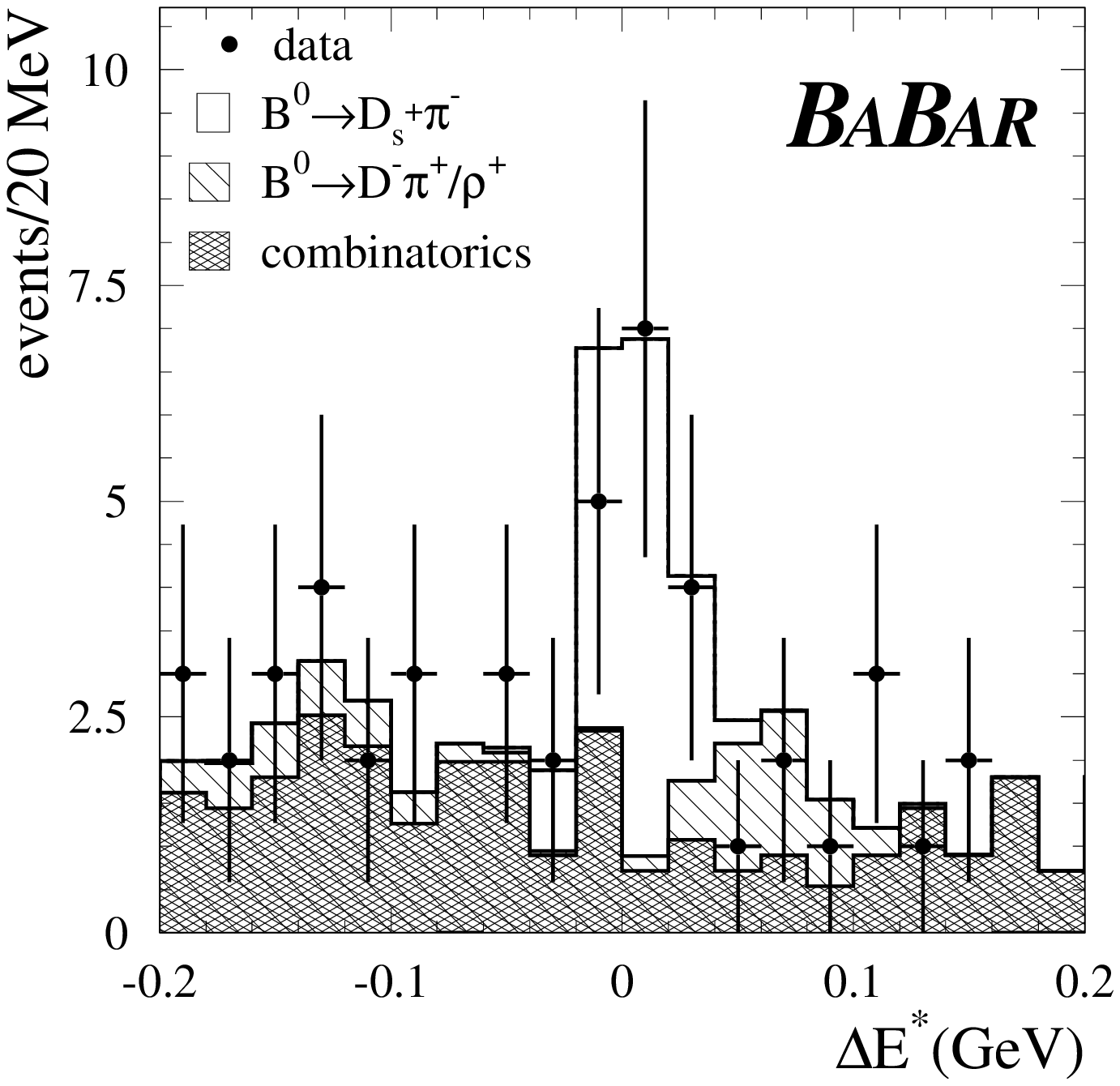}
\includegraphics[width=0.45\linewidth]{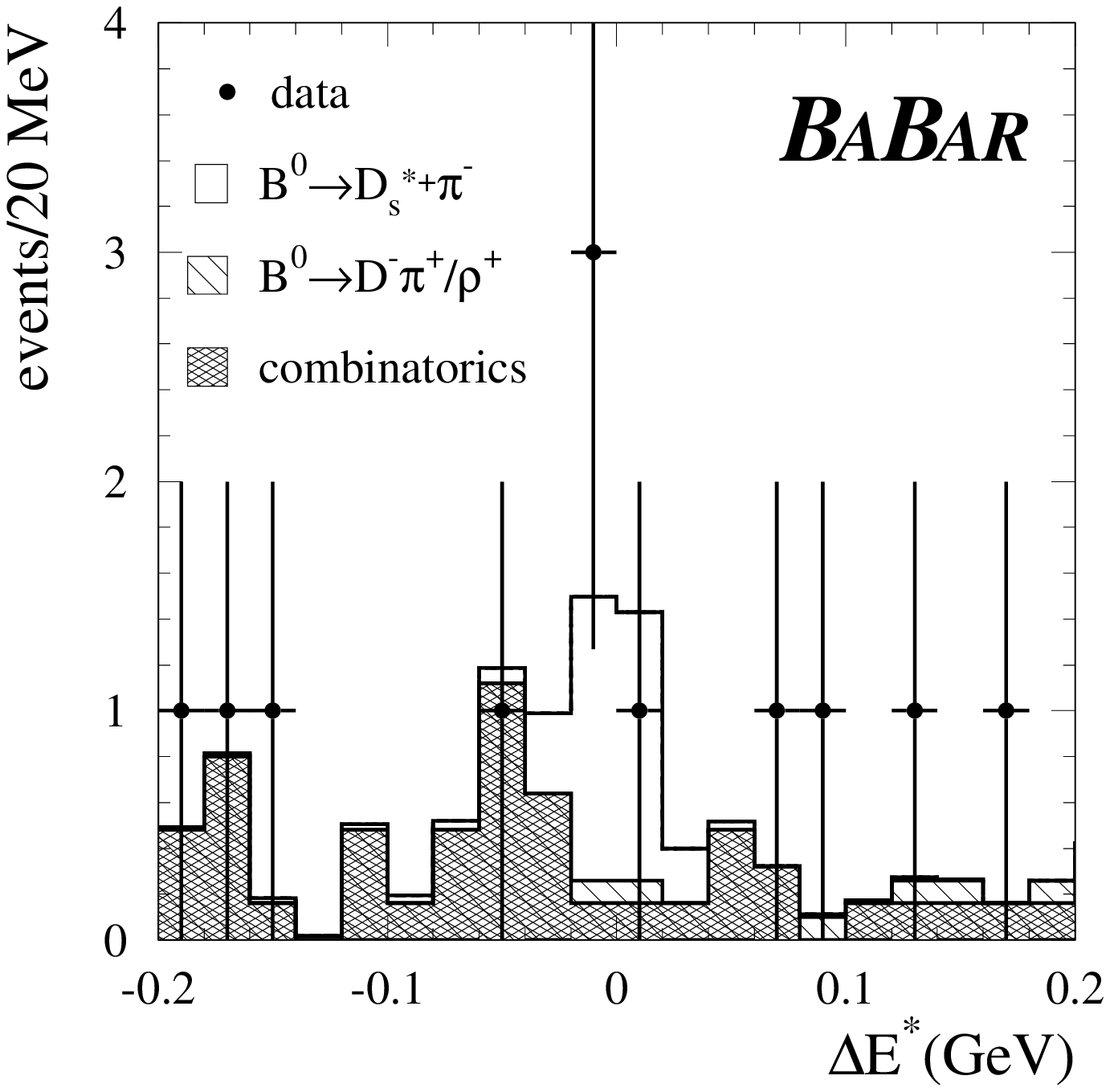}

\caption{ \de\ projection for \bdspi\  (left) and  \bdsspi\  (right) events in a 3$\sigma$ \mes\ window 
around the $B$ mass. The  signal from the Monte Carlo simulation (open histogram), 
the combinatorial background (cross hatched histogram), and the $\Bz\to \Dm\pip$ and 
$\Bz\to \Dm\rho^+$ components (hatched histogram) are overlaid. 
}\label{fig:datade}
\end{center}
\end{figure}

As a consistency check for the \bdspi\ search, we plot in Figure~\ref{fig:datade} the \de\ projection
for the $\Ds \pim$ and the $\Dss \pim$  modes after requiring  \mes\ lie within a 3$\sigma$ window of the known $B$ mass.
A comparison of the observed \de\ distribution with the 
expectations from the combinatorial background, the component of the peaking background from  $\Bz\to \Dm\pip$ and 
$\Bz\to \Dm\rho^+$,  and the signal itself shows good agreement. The other peaking components with the same final state as 
the signal are included in the signal.

The total systematic error  is the sum in quadrature  of the 
contributions shown in Table~\ref{tab:systematic}. 
The systematic uncertainty on the combinatorial background subtraction derives from
varying the background ARGUS shape within the statistical uncertainty on its determination. 
The uncertainty on the peaking background accounts for the limited size of the sample used to estimate it.
The  uncertainties on the \Ds\ and \Dss\ branching fractions are taken from~\cite{PDG2000}:
the dominant source is  \BR (\Ds$\rightarrow\phi\pip$), 
a 25\% relative error correlated among all modes, since the other branching fractions are measured relative to 
it.
The uncertainty due to the possibility that the simulation does not appropriately reproduce the shape of the event selection variables
is estimated by comparing the corresponding distributions between signal simulation and a copious and pure 
 $\Bz\to\Dstarm\pip$ control sample. 
The tracking efficiency is computed from a sample of $e^+e^-\rightarrow\tau^+\tau^-$ events, with one $\tau$ decaying into
three tracks  and one neutrino, and one decaying into one track and one neutrino.
We estimate the \KS\ efficiency uncertainty
by comparing the momentum and flight-distance distributions in data and
Monte Carlo simulation. The kaon identification efficiency  is derived from a sample of 
$D^{*+} \to D^{0}\pip,D^{0} \to \Km\pip$ decays. 

\begin{table}[!htb]
\caption{The systematic uncertainties in the measurement of
the branching fraction \BR.}
\begin{center}
\begin{tabular}{lcc}  \hline \hline
                  & \multicolumn{2}{c}{ Uncertainty ($\times 10^{-5}$)  } \\ 
                                        & \Ds\pim   &  \Dss\pim\\  \hline

\Ds\ and \Dss\ branching fractions     &0.82 & 0.53\\ 
Peaking Background  & 0.44& 0.09 \\           
Selection variables    & 0.30& 0.18\\                               
Tracking and \KS\ efficiency  & 0.17& 0.11\\           
Kaon identification       & 0.14 &0.09 \\                         
Combinatoric background                 & 0.09& 0.06\\ 
Simulation statistics         & 0.05& 0.07\\                        
$N_{BB}$          & 0.04& 0.02\\                       
Photon efficiency      & - &  0.03 \\                         \hline 
Total                 & 1.01& 0.59\\                        \hline                
\end{tabular}
\end{center}
\label{tab:systematic}
\end{table}

 
In conclusion, we observe 17 \bdspi\ and 4 \bdsspi\ candidates in the signal region.
We therefore report a 3.5$\sigma$ signal for the \btou\ transition
 \bdspi, with \brds. Given that the dominant uncertainty 
comes from the knowledge of the \Ds\ branching fractions we also compute
$\BR( \bdspi )\times \BR(\Ds\rightarrow\phi\pip)=(1.11\pm 0.37\pm 0.22) \times 10^{-6}$.
The search for \bdsspi\ yields a result of 2.2$\sigma$ significance,
 and a 68\% confidence interval \brdss, which can be translated into an upper limit \brdsslim

We are grateful for the 
extraordinary contributions of our \pep2\ colleagues in
achieving the excellent luminosity and machine conditions
that have made this work possible.
The success of this project also relies critically on the 
expertise and dedication of the computing organizations that 
support \babar.
The collaborating institutions wish to thank 
SLAC for its support and the kind hospitality extended to them. 
This work is supported by the
US Department of Energy
and National Science Foundation, the
Natural Sciences and Engineering Research Council (Canada),
Institute of High Energy Physics (China), the
Commissariat \`a l'Energie Atomique and
Institut National de Physique Nucl\'eaire et de Physique des Particules
(France), the
Bundesministerium f\"ur Bildung und Forschung
(Germany), the
Istituto Nazionale di Fisica Nucleare (Italy),
the Research Council of Norway, the
Ministry of Science and Technology of the Russian Federation, and the
Particle Physics and Astronomy Research Council (United Kingdom). 
Individuals have received support from 
the A. P. Sloan Foundation, 
the Research Corporation,
and the Alexander von Humboldt Foundation.

\end{document}